\documentclass[journal]{IEEEtran}
\usepackage[pdftex]{graphicx} 
\usepackage{amsmath} 
\usepackage{amsfonts} 
\usepackage{lipsum} 
\usepackage{booktabs}

\usepackage{bm}
\newcommand*{\B}[1]{\ifmmode\bm{#1}\else\textbf{#1}\fi}

\ifCLASSOPTIONcompsoc
  \usepackage[caption=false,font=normalsize,labelfont=sf,textfont=sf]{subfig}
\else
  \usepackage[caption=false,font=footnotesize]{subfig}
\fi

\usepackage{acronym}
\acrodef{NMF}{non-negative matrix factorization}
\acrodef{RNN}{recurrent neural network}
\acrodef{PIT}{permutation invariant training}
\acrodef{DC}{Deep Clustering}
\acrodef{DNN}{deep neural network}
\acrodef{DPRNN}{dual-path recurrent neural network}
\acrodef{TCN}{temporal convolution network}
\acrodef{SIR}{signal-to-interference ratio}
\acrodef{SI-SDR}{scale-invariant source-to-distortion ratio}
\acrodef{SDR}{source-to-distortion ratio}
\acrodef{SNR}{signal-to-noise ratio}
\acrodef{TV}{time-varying}
\acrodef{TI}{time-invariant}
\acrodef{gln}[gLN]{Global Layer Normalization}
\acrodef{ERLE}{echo return loss enhancement}
\acrodef{LSTM}{long short-term memory}
\acrodef{ESTOI}{extended short-time objective intelligibility}
\acrodef{TF}{time-frequency}
\acrodef{ReLU}{rectified linear unit}

\acrodef{SS}{speaker separation}
\acrodef{SE}{speaker extraction}
\acrodef{SV}{speaker verification}

\acrodef{CI}{complete information}
\acrodef{DI}{distorted information}

\newcommand{\SEA}{SE\mbox{-}A}
\newcommand{\SEV}{SE\mbox{-}V}
\newcommand{\TCD}{TCD\mbox{-}TIMIT}

\usepackage[framemethod=tikz]{mdframed}

\usepackage{cite}

\usepackage{orcidlink}

\usepackage{hyperref}   
\hypersetup{hidelinks}                     %

\usepackage{multirow}

\usepackage[tight-spacing=true]{siunitx} 

\usepackage{verbatim} 

\begin{document}

\title{New Insights on Target Speaker Extraction}

\author{Mohamed~Elminshawi,~\IEEEmembership{Student Member,~IEEE,}
        Wolfgang~Mack,~\IEEEmembership{Student Member,~IEEE,}
        Srikanth~Raj~Chetupalli,~\IEEEmembership{Member,~IEEE,}
        Soumitro~Chakrabarty,~\IEEEmembership{Member,~IEEE,}
        and~Emanu\"{e}l~A.~P.~Habets,~\IEEEmembership{Senior Member,~IEEE}
        
\thanks{The authors are with the International Audio Laboratories Erlangen (a joint institution of the Friedrich Alexander University Erlangen-Nürnberg (FAU) and Fraunhofer IIS), 91058 Erlangen, Germany (\mbox{e-mail}: \href{mailto:mohamed.elminshawi@audiolabs-erlangen.de}{\nolinkurl{mohamed.elminshawi@audiolabs-erlangen.de}}, \href{mailto:wolfgang.mack@audiolabs-erlangen.de}{\nolinkurl{wolfgang.mack@audiolabs-erlangen.de}}, \href{mailto:srikanth.chetupalli@iis.fraunhofer.de}{\nolinkurl{srikanth.chetupalli@iis.fraunhofer.de}}, 
\href{mailto:soumitro.chakrabarty@audiolabs-erlangen.de}{\nolinkurl{soumitro.chakrabarty@audiolabs-erlangen.de}}, \href{mailto:emanuel.habets@audiolabs-erlangen.de}{\nolinkurl{emanuel.habets@audiolabs-erlangen.de}}).}
}



\maketitle

\begin{abstract}
\Ac{SE} aims to segregate the speech of a target speaker from a mixture of interfering speakers with the help of auxiliary information. Several forms of auxiliary information have been employed in single-channel \ac{SE}, such as a speech snippet enrolled from the target speaker or visual information corresponding to the spoken utterance. {\color{black}The effectiveness of the auxiliary information in \ac{SE} is typically evaluated by comparing the extraction performance of \ac{SE} with uninformed \ac{SS} methods. Following this evaluation protocol, many \ac{SE} studies have reported performance improvement compared to \ac{SS}, attributing this to the auxiliary information.} However, such studies have been conducted on a few datasets and have not considered {\color{black} recent \acl{DNN} architectures} for \ac{SS} that have shown impressive separation performance. In this paper, we examine the role of the auxiliary information in \ac{SE} for different input scenarios and over multiple datasets. Specifically, we compare the performance of two \ac{SE} systems (audio-based and video-based) with \ac{SS} using a common framework that utilizes the {\color{black} recently proposed} dual-path recurrent neural network as the main learning machine. {\color{black}Experimental evaluation on various datasets demonstrates that the use of auxiliary information in the considered \ac{SE} systems does not always lead to better extraction performance compared to the uninformed \ac{SS} system. Furthermore, we offer insights into the behavior of the \ac{SE} systems when provided with different and distorted auxiliary information given the same mixture input.} 

\end{abstract}

\acresetall

\begin{IEEEkeywords}
Speaker extraction, speaker separation, deep learning, single-channel.
\end{IEEEkeywords}

%
\IEEEpeerreviewmaketitle

\section{Introduction}
\IEEEPARstart{S}{peech} is the primary means through which humans communicate. In typical acoustic scenes, a recording of a speaker of interest is often degraded by other acoustic sources, such as background noise and interfering speakers. Remarkably, human brains have the ability to focus on a specific acoustic source in a noisy environment while ignoring other sources, a phenomenon commonly referred to as the \textit{cocktail party effect} \cite{cherry1953some}. In contrast, speech corrupted with concurrent interfering speakers has been shown to severely deteriorate the performance of several speech processing algorithms, including automatic speech recognition~\cite{cooke2010monaural} and \ac{SV}~\cite{martin2001speaker}. Over the past several decades, a considerable amount of research has been devoted to dealing with overlapping speech as a \acfi{SS} task, i.e., separating all speakers from the observed mixture signal~\cite{wang2018supervised}. In particular, deep learning has considerably advanced the performance of single-channel \ac{SS} methods \cite{hershey2016deep, yu2017permutation, chen2017deep, Luo2019, Luo2019a, chen2020dual, zeghidour2021wavesplit, subakan2021attention, byun2021monaural}. One fundamental problem associated with \ac{SS} is the permutation problem, i.e., the correspondence between the separated output signals and the speakers is arbitrary. This ambiguity poses a challenge when training \acp{DNN} for separation since the loss function needs to be computed between each output signal and the ground-truth speech of its corresponding speaker. {\color{black} To address this challenge, \ac{PIT} \cite{yu2017permutation, kolbaek2017multitalker} has been proposed, which enables optimizing \acp{DNN} that directly separate the speech signals by finding the permutation of the ground-truth signals that best matches the output signals.}


In many scenarios, it may not be necessary to reconstruct all speakers from the mixture; instead, it suffices to extract a single target speaker. This task has been given numerous names in the literature, among which are \textit{target speaker extraction} \cite{delcroix2018single, delcroix2021speaker}, \textit{informed speaker extraction} \cite{ochiai2019unified}, or simply \acfi{SE} \cite{xu2020spex, vzmolikova2019speakerbeam}. In contrast to \ac{SS}, \ac{SE} systems do not suffer from the permutation ambiguity since only {\color{black}a} single output exists. Early works on speaker extraction \cite{du2014speech, du2016regression} were target-dependent, i.e., systems designed to extract speech from only a particular speaker and cannot generalize to other speakers. Such systems require abundant training data from the target speaker, which is infeasible in many applications. To realize speaker-independent \ac{SE} systems, prior knowledge or auxiliary information must be provided to specify the target signal. \ac{SE} approaches can be categorized based on the modality of the auxiliary information. Audio-based \ac{SE} (\SEA{}) methods rely on a speech snippet from the target speaker that guides the system towards that speaker \cite{vzmolikova2019speakerbeam, wang2018voicefilter, wang2018deep, Delcroix2020}. Visual-based \ac{SE} (\SEV{}) methods\footnote{Also known in the literature as audio-visual speaker enhancement/separation methods.} have also been proposed that leverage visual information from the target speaker, such as lip movements \cite{gabbay2017visual, hou2018audio, afouras2018conversation, wu2019time} or cropped facial frames \cite{ephrat2018looking, afouras2019my}. Other methods have exploited multi-modal information, for example, by utilizing both visual features of the target speaker as well as an enrollment utterance \cite{afouras2019my, Ochiai2019}. Finally, brain signals \cite{ceolini2020brain} and speaker activity \cite{delcroix2021speaker} have also been utilized as auxiliary signals for \ac{SE}. 

{\color{black} Clearly, \ac{SS} and \ac{SE} are related problems in the sense that both deal with overlapped speech. In fact, \ac{SE} can be realized by first separating all speakers using a \ac{SS} system followed by \ac{SV} for target speaker selection. However, \ac{SS} and \ac{SE} exhibit notable distinctions in terms of their underlying assumptions and the nature of errors that could arise. In \ac{SS}, all speakers in the mixture are to be recovered, whereas only a unique speaker is assumed to be the target in \ac{SE}. In addition, knowledge about the number of speakers in the mixture is often assumed in \ac{SS} \cite{hershey2016deep, isik2016single, yu2017permutation, kolbaek2017multitalker}, while such an assumption is not necessary in \ac{SE}. Furthermore, \ac{SE} necessitates prior knowledge about the target speaker in the form of an auxiliary signal, while \ac{SS} blindly segregates the speech signals. With respect to evaluation, any permutation of the outputs of SS is a valid solution and leads to the same objective metrics. In contrast, SE systems are prone to \emph{speaker confusion}, i.e., recovering an interfering speaker instead of the target \cite{zhao22b_interspeech}. The above points should not be overlooked when evaluating and, especially, comparing the performance of SS and SE.}

{\color{black}The utility of the auxiliary information in \ac{SE} is generally assessed by comparing the extraction performance of \ac{SE} with \ac{SS} following an oracle evaluation protocol}, i.e., each output of the \ac{SS} system is compared with the ground-truth speech of the target speaker, and the best match is selected as the estimated target. Using this protocol, the majority of \ac{SE} works often report performance improvement over \ac{SS}, arguing that this is due to the use of auxiliary information. In particular, it has been argued that the use of auxiliary information improves the performance in scenarios involving mixtures having similar voice characteristics (e.g., same-gender mixtures)~\cite{gabbay2017visual}, long mixtures with complicated overlapping patterns~\cite{vzmolikova2019speakerbeam}, or adverse acoustic conditions, e.g., very low \acp{SNR} or more interfering speakers \cite{chuang2020lite, michelsanti2021overview}. Another work in \SEV{} has demonstrated that the auxiliary visual information improves the extraction performance compared to \ac{SS}, especially for visually distinguishable sounds \cite{aldeneh2021role}. However, such studies have been conducted on a few datasets and have not considered {\color{black} more recently proposed} \ac{DNN} architectures that have achieved unprecedented performance in \ac{SS} \cite{Luo2019a, chen2020dual, subakan2021attention}. 


It is, thus, imperative to validate the effectiveness of the auxiliary information in \ac{SE}. Having a clear understanding of the contribution of the auxiliary information in \ac{SE} would not only give us more insights into how such systems function, but it could also allow us to develop robust \ac{SE} systems against unreliable auxiliary information, e.g., noisy enrollment utterances, occluded or temporally misaligned visual features. In this work, we objectively examine the role of the auxiliary information in \ac{SE}. Specifically, the performance of \SEA{} and \SEV{} systems is compared to \ac{SS} across several datasets. To establish a fair comparison, we construct a common framework for all systems that employs the {\color{black} recently proposed} \ac{DPRNN} architecture \cite{Luo2019a} as the main learning machine. {\color{black}In addition, we provide insights into the role of the auxiliary information in \ac{SE} by inspecting the behavior of the systems for different and distorted auxiliary information given the same mixture signals.}

The remainder of the paper is organized as follows. Section~\ref{sec: problem_formulation} describes the tasks of \ac{SS} and \ac{SE} as well as the systems employed in this study. In Section~\ref{sec: experimental_design}, the experimental setup is discussed, and Section~\ref{sec: experimental_results} presents the experimental results. Finally, the discussion is provided in Section~\ref{sec: discussion}.

\section{Speaker Separation and Extraction Systems}
\label{sec: problem_formulation}

In this section, we formally define the problems of \ac{SS} and \ac{SE}, and provide a detailed description of the different systems used in this study, which resemble{\color{black}, to a great extent,} {\color{black} recently developed} methods in the literature on \ac{SS} and \ac{SE} \cite{Luo2019, Luo2019a, Delcroix2020, wu2019time, Ochiai2019}. Figure~\ref{fig: pipelines} shows the block diagrams of the different systems. {\color{black} Note that a common backbone is employed for all systems to ensure a fair comparison.} 
Further details about the systems' configurations are described in Section~\ref{subsec:model_configurations}. 

Let $\bm{y} \in \mathbb{R}^S$ be $S$ samples of an observed single-channel time-domain mixture signal consisting of speech from $C$ speakers, denoted by $\bm{x}_1, \ldots, \bm{x}_C \in \mathbb{R}^S$, i.e., 
\begin{equation}
	\bm{y} = \sum_{i=1}^{C} \bm{x}_i.
\end{equation} 

\subsection{Speaker Separation (\acs{SS})}
\label{speaker_separation}

\begin{figure*}[t]
\centering
\subfloat[\ac{SS}]{\includegraphics[height=3in]{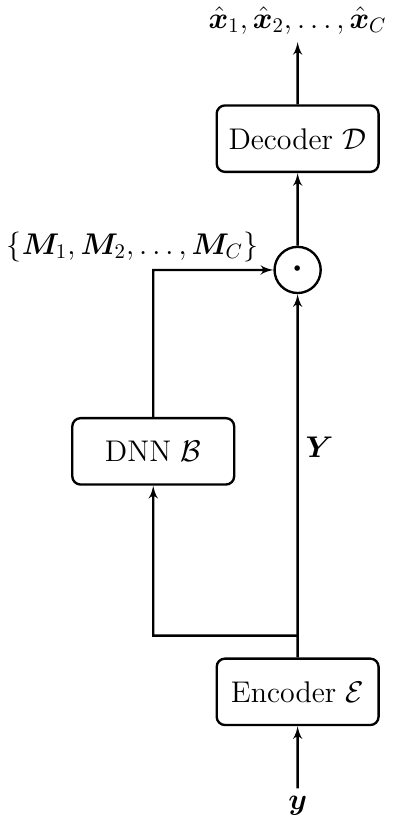}%
\label{fig: audio_only}}
\hfil
\subfloat[\SEA{}]{\includegraphics[height=3in]{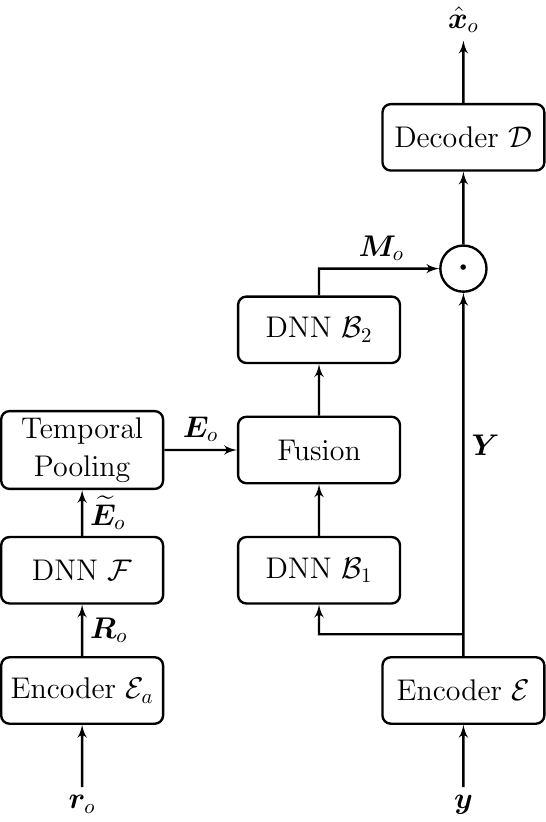}%
\label{fig: audio_ref}}
\hfil
\subfloat[\SEV{}]{\includegraphics[height=3in]{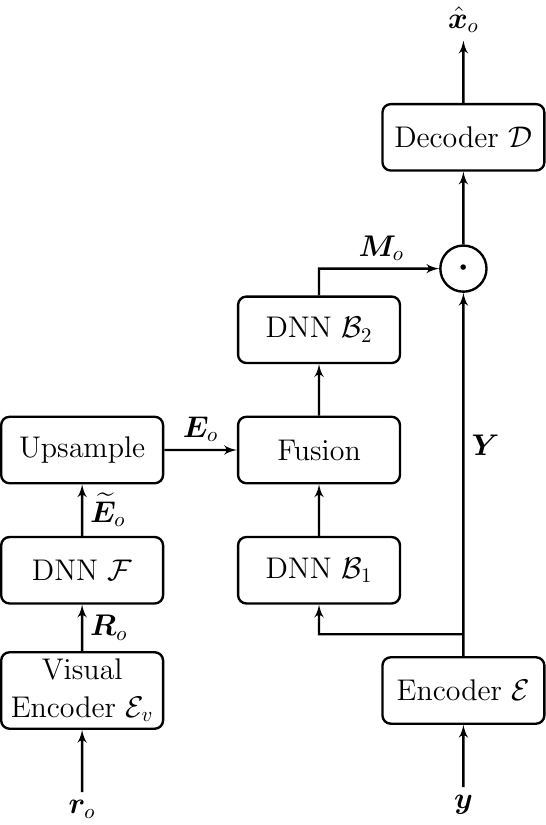}%
\label{fig: video_ref}}
\caption{Block diagrams of the \acf{SS} and \acf{SE} systems: (a) \ac{SS} trained with u\ac{PIT}, (b) audio-based \ac{SE} (\SEA{}), and (c) video-based \ac{SE} (\SEV{}). }
\label{fig: pipelines}
\end{figure*}

The objective of \a{SS} is to reconstruct all the constituent speech signals in the mixture, i.e.,
\begin{equation}
	\{\hat{\bm{x}}_1, \hat{\bm{x}}_2, \ldots , \hat{\bm{x}}_C\} =  f(\bm{y}),
\end{equation} 

\noindent
where $\hat{\bm{x}}_i \in \mathbb{R}^S$ denotes the estimated speech signal at the $i$-th output, and $f$ represents the transformation applied by the separation system on the mixture signal $\bm{y}$. Note that the order of the output signals is arbitrary, and a mapping between each output and its corresponding speaker identity is typically required. Following state-of-the-art \a{SS} methods \cite{Luo2018, Luo2019, chen2020dual, subakan2021attention}, we adopt an encoder-masker-decoder structure for $f$, as shown in Figure~\ref{fig: audio_only}. In particular, $f$ comprises three main blocks: an encoder, a mask estimator using a \ac{DNN}, and a decoder, represented by $\mathcal{E}$, $\mathcal{B}$, and $\mathcal{D}$, respectively. The encoder $\mathcal{E}$ transforms the time-domain waveform of the mixture into frame-wise features $\bm{Y} \in \mathbb{R}^{N \times T}$, where $N$ is the dimensionality of the encoded features of each time frame, and $T$ denotes the number of time frames, i.e.,
\begin{equation}
	\bm{Y} =   \mathcal{E}(\bm{y}).
	\label{encoder_audio}
\end{equation} 

\noindent
The mask estimator $\mathcal{B}$ is a \ac{DNN} that maps the encoded features $\bm{Y}$ to a mask for each speaker in the mixture, i.e.,
\begin{equation}
	\{\bm{M}_1, \bm{M}_2, \ldots, \bm{M}_C\} =  \mathcal{B}(\bm{Y}),
\end{equation} 

\noindent
where $\bm{M}_i \in \mathbb{R}^{N \times T}$ denotes the mask for the $i$-th output. Each mask is then applied to the encoded features $\bm{Y}$ and subsequently fed to the decoder $\mathcal{D}$ to reconstruct the time-domain waveform of the corresponding speaker, denoted by $\hat{\bm{x}}_i \in \mathbb{R}^S$, as
\begin{equation}
	\hat{\bm{x}}_i =  \mathcal{D}(\bm{Y} \odot \bm{M}_i),
\end{equation} 

\noindent
where $\odot$ denotes the Hadamard product. To optimize the parameters of the separation system, utterance-level \ac{PIT} (u\ac{PIT}) \cite{kolbaek2017multitalker} is used first to find the bijective mapping between each output signal $\hat{\bm{x}}_i$ and its corresponding speaker, as
\begin{equation}
	\phi^* = \underset{\phi \in \mathcal{P}}{\operatorfont{arg\,min}} \sum_{i=1}^{C} \bm{\ell}(\bm{x}_{\phi (i)}, \hat{\bm{x}}_i),
\end{equation}

\noindent
where $\bm{\ell}$ is a {\color{black} loss function defined} between two time-domain signals, $\mathcal{P}$ represents the set of all possible permutations, and $\phi^*$ is the optimum permutation that provides the minimum loss. The {\color{black} total loss} $\mathcal{L}$ is then computed, as
\begin{equation}
	\mathcal{L} = \sum_{i=1}^{C} \bm{\ell}(\bm{x}_{\phi^* (i)}, \hat{\bm{x}}_i).
\end{equation}


\subsection{Speaker Extraction (\acs{SE})}

In contrast to \ac{SS}, \ac{SE} refers to the task of reconstructing a single target speaker from the mixture given auxiliary information about the target speaker. 
We refer to the auxiliary information as the reference signal and denote it by $\bm{r}_o$, where $o$ is the index of the desired speaker. Note that the dimensionality of $\bm{r}_o$ depends on the type of information provided to the system. \Ac{SE} can be formulated as
\begin{equation}
	\hat{\bm{x}}_o =  g(\bm{y}, \bm{r}_o),
\end{equation} 

\noindent
where $g$ represents the transformation carried out by the \ac{SE} system, and $\hat{\bm{x}}_o \in \mathbb{R}^S$ denotes the estimated speech of the target speaker. State-of-the-art \ac{SE} methods decompose $g$ into two stages \cite{ephrat2018looking, wu2019time, Ochiai2019, wang2018voicefilter, Delcroix2020}: an auxiliary network and an extraction network, represented by $h$ and $\tilde{f}$, respectively. The auxiliary network $h$ extracts informative features, denoted by $\bm{E}_o$, from the reference signal, which help to specify the target speaker, i.e.,
\begin{equation}
	\bm{E}_o =  h(\bm{r}_o).
		\label{aux_network}
\end{equation} 

\noindent
The second stage is to condition the extraction network $\tilde{f}$ on the features $\bm{E}_o$ such that an estimate of the target speaker can be obtained as
\begin{equation}
	\hat{\bm{x}}_o =  \tilde{f}(\bm{y}, \bm{E}_o).
\end{equation} 

\noindent
{\color{black}Following prior works \cite{wu2019time, Delcroix2020},} $\tilde{f}$ has a similar encoder-masker-decoder structure to the separation system $f$ described in Section \ref{speaker_separation}, except that the mask estimator consists of two \ac{DNN} blocks $\mathcal{B}_1$ and $\mathcal{B}_2$, as well as a fusion layer inserted in between, such that the informative features $\bm{E}_o$ can be included. Many fusion techniques have been proposed, e.g., concatenation-based \cite{ephrat2018looking, wu2019time}, and product-based \cite{Ochiai2019, vzmolikova2019speakerbeam, Delcroix2020}. In this work, we adopt the latter technique as the fusion mechanism, similar to \cite{vzmolikova2019speakerbeam, Ochiai2019}. The transformations carried out by $\tilde{f}$ are represented by
\begin{equation}
	\bm{M}_o = \mathcal{B}_2(\mathcal{B}_1(\bm{Y}) \odot \bm{E}_o),
	\label{fusion}
\end{equation} 

\begin{equation}
	\hat{\bm{x}}_o =  \mathcal{D}(\bm{Y} \odot \bm{M}_o).
\end{equation} 

\noindent
The loss function, in this case, is computed with respect to the target speaker only, and hence u\ac{PIT} is not required, i.e.,
\begin{equation}
	\mathcal{L} = \bm{\ell}(\bm{x}_o, \hat{\bm{x}}_o).
\end{equation}

\noindent
The design of the auxiliary network $h$ in (\ref{aux_network}) depends on the modality of the reference signal $\bm{r}_o$. In this work, we focus on audio-based \ac{SE} (\SEA{}) and video-based \ac{SE} (\SEV{}). Both systems are illustrated in Figure~\ref{fig: audio_ref} and Figure~\ref{fig: video_ref}, and described in the following. 


\subsubsection{Audio-based \ac{SE} (\SEA{})}
\label{sec: audio_based}
A reference speech signal from the target speaker {\color{black}is} used as auxiliary information to guide the extraction system. Typical \SEA{} methods realize this process by mapping the reference speech signal to an embedding vector that encodes the voice characteristics of the target speaker. The well-known speaker representations developed for speaker recognition, such as \mbox{i-vector}~\cite{dehak2010front} and \mbox{d-vector}~\cite{wan2018generalized}, have been employed in \ac{SE} in \cite{vzmolikova2019speakerbeam} and \cite{wang2018voicefilter}. Alternatively, speaker representations can also be learned via an auxiliary \ac{DNN} jointly optimized with the extraction system \cite{vzmolikova2019speakerbeam, Delcroix2020}. 

In this study, we adopt the end-to-end approach of training the auxiliary network jointly with the extraction network. The first block in the auxiliary network is an audio encoder $\mathcal{E}_a$ (similar to $\mathcal{E}$ in (\ref{encoder_audio})), i.e.,
\begin{equation}
	\bm{R}_o = \mathcal{E}_a(\bm{r}_o),
\end{equation} 

\noindent
where $\bm{r}_o \in \mathbb{R}^{S_r}$ denotes the speech reference signal from the target speaker having a length of $S_r$ samples, and \mbox{$\bm{R}_o \in \mathbb{R}^{N \times T_{a}}$} represents the encoded frame-wise features, where $N$ is the dimensionality of the features of each time frame, and $T_{a}$ represents the number of time frames. The features $\bm{R}_o$ are further processed with a \ac{DNN}, denoted by $\mathcal{F}$, which produces an embedding vector for each time frame, arranged in the matrix $\widetilde{\bm{E}}_o \in \mathbb{R}^{N \times T_{a}}$, as
\begin{equation}
	\widetilde{\bm{E}}_o =  \mathcal{F}(\bm{R}_o).
\end{equation}

\noindent
Temporal average pooling is {\color{black} then} applied {\color{black} to} the frame-wise features $\widetilde{\bm{E}}_o$, such that an utterance-wise embedding vector, represented by $\bm{E}_o \in \mathbb{R}^{N \times 1}$, is obtained. Note that, in this case, the embedding vector $\bm{E}_o$ is \textit{time-invariant}, and it is broadcasted over the different time frames in the fusion layer in (\ref{fusion}). 

\subsubsection{Video-based \ac{SE} (\SEV{})}
\label{sec: video_based}
 In an attempt to mimic the multimodality of human perception \cite{golumbic2013visual, partan1999communication}, video-based \ac{SE} leverages visual cues, such as lip movements or facial expressions, as auxiliary information. The first component of the auxiliary network in the \SEV{} system is a visual encoder, denoted by $\mathcal{E}_v$, that extracts visual features $\bm{R}_o \in \mathbb{R}^{N_v \times T_{v}}$, where $N_v$ represents the dimensionality of the features, and $T_{v}$ represents the number of time frames, from the given reference visual signal $\bm{r}_o \in \mathbb{R}^{D \times H \times W \times T_{v}}$, where $D$, $H$, and $W$ denotes the depth, height, and width, respectively, i.e., 
\begin{equation}
	\bm{R}_o =  \mathcal{E}_v(\bm{r}_o).
\end{equation} 

\noindent
The design of the visual encoder $\mathcal{E}_v$ depends on the type of visual information used. When facial frames are utilized as a reference signal, typically $\mathcal{E}_v$ is a pre-trained face recognition network, e.g., FaceNet \cite{cole2017synthesizing}, from which an embedding vector for each facial frame is extracted \cite{ephrat2018looking, Ochiai2019}. In the case of lip frames as a reference signal, $\mathcal{E}_v$ typically consists of a \mbox{spatio-temporal} convolutional layer, i.e., \mbox{3-D~ConvLayer}, followed by \mbox{ResNet-18} \cite{stafylakis2017combining} that outputs a lip embedding for each frame \cite{afouras2018conversation, wu2019time}. It has been shown that using lip features as visual information generally provides better extraction performance than facial features \cite{inan122019evaluating, shetu2021empirical}. 

In this study, we adopt lip frames as visual information and use the \mbox{3-D~ConvLayer} + \mbox{ResNet-18} structure for the visual encoder $\mathcal{E}_v$. The {\color{black}visual} embeddings $\bm{R}_o$ are further processed with a \ac{DNN} $\mathcal{F}$, resulting in more {\color{black}task-specific} features, represented by $\widetilde{\bm{E}}_o \in \mathbb{R}^{N \times T_{v}}$, as
\begin{equation}
	\widetilde{\bm{E}}_o =  \mathcal{F}(\bm{R}_o).
\end{equation}

\noindent
Finally, to match the sampling rates of the visual and audio streams, the learned features $\widetilde{\bm{E}}_o$ are upsampled using linear interpolation along the temporal dimension, similar to \cite{Owens_2018_ECCV, pan2022usev}, resulting in the frame-wise features $\bm{E}_o \in \mathbb{R}^{N \times T}$.

\section{Experimental Design}
\label{sec: experimental_design}

In this section, we introduce the datasets used in this study, followed by the specific configurations of the systems presented in Section~\ref{sec: problem_formulation}. {\color{black} Finally, the training setup and the evaluation protocol are described.} 

\subsection{Datasets}
\label{sec: datasets}
{\color{black}
For experimentation, we consider the four most commonly used datasets in the literature on \ac{SE}. These datasets differ in terms of the dataset size, the number of speaker identities, the vocabulary size, and the recording conditions. The details of these datasets are presented in Table~\ref{tab: datasets}. The audio files in all datasets have a sampling frequency of 16~kHz.


\subsubsection{TCD-TIMIT \cite{harte2015tcd}} This corpus consists of synchronized audio-visual recordings of 59 speakers reading sentences from the TIMIT corpus. TCD-TIMIT is collected in a controlled environment, thus comprising high-quality audio and video clips of speech. The video recordings are sampled at 25~frames~per~second.

Since there are no official \ac{SS}/\ac{SE} dataset splits for \TCD{}, we created training, validation, and test splits based on speaker identities, i.e., the splits are ensured to form disjoint sets in terms of the speaker identities. We then simulated \mbox{2-speaker} and \mbox{3-speaker} mixtures by randomly sampling utterances from different speakers in each split and mixing them with a \ac{SIR} sampled from $-$5~dB to 5~dB. The mixtures have a duration of 3~seconds. In the case of \mbox{3-speaker} mixtures, two \acp{SIR} were sampled, and each interferer was scaled by its corresponding \ac{SIR} with respect to the target, and then all signals were superimposed. 



\subsubsection{LRS3 \cite{afouras2018lrs3}} It is a large-scale audio-visual corpus obtained from TED and TEDx talks. Unlike \TCD{}, LRS3 is collected \textit{in the wild}, resulting in a lower quality of samples compared to \TCD{}. However, LRS3 has a tremendous variability in terms of the spoken sentences, visual appearances, and speaking styles, which allows developing robust \ac{DNN} models that generalize to real-world conditions. Similar to \TCD{}, the video recordings are sampled at 25~frames~per~second. In addition, we followed the same procedure as in \TCD{} to create dataset splits suitable for \ac{SS}/\ac{SE}, since there are no official splits for LRS3. 

\subsubsection{WSJ0Mix \cite{hershey2016deep}} Created from the WSJ0 corpus \cite{garofolo1993csr}, the popular WSJ0-2Mix and WSJ0-3Mix datasets for single-channel 2~and~3~speaker mixtures, respectively, have become the standard benchmark for SS. 
The utterances are mixed with a \ac{SIR} randomly sampled from $-$5~dB to 5~dB. We used the \emph{min} version of the datasets. Unlike \TCD{} and LRS3, WSJ0Mix comprises only audio signals, and there are no corresponding visual recordings of the speakers. 

\subsubsection{LibriMix \cite{cosentino2020librimix}} It is an audio-only dataset that comprises \mbox{2-speaker} and \mbox{3-speaker} mixtures created from the \mbox{LibriSpeech} corpus \cite{Libri2015}. The \ac{SIR} of the mixtures in the dataset follows a normal distribution with a mean of 0~dB and a standard deviation of 4.1~dB. Similar to WSJ0Mix, we used the \emph{min} version of the dataset.

For the above datasets, the reference speech signal for the \SEA{} system was an utterance spoken by the target speaker that was different from the one in the mixture. For the \SEV{} system, the reference signal was obtained by cropping the lip region of the target speaker in the video clip corresponding to the spoken utterance. 



}




\begin{table}[t]
\centering
\renewcommand{\arraystretch}{1.7}
\caption{Statistics of the training, validation, test splits of the different datasets.}
\resizebox{\columnwidth}{!}{
\begin{tabular}{@{}lccc@{}}
\toprule
\textbf{Corpus} & \textbf{No. Speakers}  & \textbf{Total Duration [h]} & \textbf{No. Utterances} \\ \midrule
TCD-TIMIT \cite{harte2015tcd} & 47 / 6 / 6  & 6.7 / 0.9 / 0.8 & 30k / 5k / 3k \\
LRS3 \cite{afouras2018lrs3} & 906 / 49 / 142 & 25.0 / 1.2 / 3.7 & 30k / 5k / 3k  \\
WSJ0 \cite{garofolo1993csr}  & 101 / (8 / 10)  & 24.9 / (1.5 / 2.2) & WSJ0-2\&3Mix \cite{hershey2016deep}: 20k / 5k / 3k \\
LibriSpeech \cite{Libri2015} & 921 / 40 / 40  & 362.4 / 5.4 / 5.4 & Libri-2Mix(3Mix) \cite{cosentino2020librimix}: 51k (34k) / 3k / 3k \\ \bottomrule
\end{tabular}
\label{tab: datasets}
}
\end{table}

\subsection{Model Configurations}
\label{subsec:model_configurations}

As mentioned in Section~\ref{sec: problem_formulation}, we adopted an encoder-masker-decoder structure for processing the input mixture signal for the different \ac{SS} and \ac{SE} systems. In particular, we followed a {\color{black} TasNet-like structure \cite{Luo2018, Luo2019}} for the encoder and decoder, which utilizes learnable kernels instead of the traditional pre-defined Fourier bases. Both audio encoders $\mathcal{E}$ and $\mathcal{E}_a$, shown in Figure~\ref{fig: pipelines}, consisted of a 1-D convolutional layer followed by a \ac{ReLU} non-linearity. We set the parameters of this layer as follows: number of kernels $N=256$, kernel size $L=32~(2~\text{ms})$, and hop size $R=16~(1~\text{ms})$. The decoder $\mathcal{D}$ was a 1-D transposed convolutional layer, having the same kernel and hop sizes as the encoder. 
{\color{black}For the \SEV{} system, the visual encoder $\mathcal{E}_v$ was pre-trained\footnote{The weights of the visual encoder is obtained from: \url{https://github.com/smeetrs/deep_avsr}} on a speech recognition task, and we kept its parameters fixed during training, similar to \cite{wu2019time}. The visual encoder is then followed by a linear layer to reduce the output dimensionality to $N_v=256$.}

For the \ac{DNN} blocks in Figure~\ref{fig: pipelines}, i.e., $\mathcal{B}$, $\mathcal{B}_1$, $\mathcal{B}_2$, and $\mathcal{F}$, we employed the {\color{black} recently proposed} \ac{DPRNN} architecture \cite{Luo2019a}. We used the \ac{DPRNN} implementation provided by SpeechBrain \cite{SB2021} with the following hyperparameters. For the intra- and inter-chunk \acp{RNN}, bi-directional \ac{LSTM} networks \cite{Hochreiter1997} were used with 128 hidden units in each direction. The bottleneck size was set to 64. We used a chunk size {\color{black} of 90, except for $\mathcal{F}$ in \SEV{}, which was set to 12. This choice ensures a comparable sequence length for the intra- and inter-chunk \acp{RNN} and is calculated for a 3-second input.} \ac{gln} \cite{Luo2019} was used, and \ac{ReLU} non-linearity was applied at the output. For the \SEA{} and \SEV{} systems, the \ac{DNN} blocks $\mathcal{B}_1$ and $\mathcal{B}_2$ were identical, and each consisted of {\color{black} 3} \ac{DPRNN} blocks, whereas for the SS system, the \ac{DNN} block $\mathcal{B}$ comprised {\color{black} 6} \ac{DPRNN} blocks. This ensures that the separation network $f$ and the extraction network $\tilde{f}$ are similar. {\color{black} In the auxiliary network of \SEA{} and \SEV{}, the \ac{DNN} block $\mathcal{F}$ consists of only one DPRNN block.}

As visual information for the \SEV{} system, lip regions were extracted using facial landmark detection implemented in \cite{king2009dlib}. The lip regions were transformed into grayscale (i.e., $D=1$) and resized to $100 \times 50$ pixels corresponding to the width $W$ and height $H$, respectively. For visual frames where the lip detection algorithm failed to detect the lip region, e.g., due to occlusion, a patch of zeros was used instead. 

{\color{black}
\subsection{Baselines}
\label{subsec:baselines}

 To validate the design choices of the adopted \ac{SE} systems, we compare their performance with two \ac{SE} baselines \cite{ge2020spex+, pan2022usev}. As a baseline for \SEA{}, we used the \mbox{SpEx+}\footnote{Official implementation provided Online: \url{https://github.com/xuchenglin28/speaker_extraction_SpEx}}~\cite{ge2020spex+}, a fully time-domain \SEA{} method comprising a multi-scale speech encoder and decoder. Following the notations in \cite{ge2020spex+}, the hyperparameters of \mbox{SpEx+} were set as follows: $L_1=40~(2.5~\text{ms})$, $L_2=160~(10~\text{ms})$, $L_3=320~(20~\text{ms})$, $N=256$, $B=8$, $R=4$, $O=256$, $P=512$, $Q=3$, $N_R=3$, $\alpha=0.1$, $\beta=0.1$, $\gamma=0.5$, and the speaker embedding dimension was set to $256$. As a baseline for \SEV{}, the recently proposed USEV\footnote{Official implementation provided Online: \url{https://github.com/zexupan/USEV}} method~\cite{pan2022usev} was used. The USEV baseline and the \SEV{} system used in this study are similar, except for the \ac{DNN} architecture used in the auxiliary network, where a temporal convolutional network (TCN) is used in USEV. Following the notations in \cite{pan2022usev}, the hyperparameters of USEV were set as follows: $L=40~(2.5~\text{ms})$, $B=64$, $N=256$, $R=6$, $H=128$, $K=100$, and 5 repeated TCN blocks were used in the auxiliary network.}


\subsection{Training Setup}
For the dataset presented in Section~\ref{sec: datasets}, we trained the \ac{SS}, \SEA{}, and \SEV{} systems as well as the baselines independently on \mbox{2-speaker} and \mbox{3-speaker} mixtures. Note that the video-based extraction systems (\SEV{} \& USEV~\cite{pan2022usev}) were not trained on WSJ0Mix and LibriMix due to the lack of visual recordings in such datasets.
Adam \cite{Kingma2015} optimizer was used with an initial learning rate of $10^{-3}$ and a weight decay of $10^{-5}$. The batch size was set to 24, and gradients were clipped if their $L_2$ norm exceeded 5. The maximum number of epochs was set to 300. A scheduler was utilized to reduce the learning rate by a factor of two if no reduction in the validation loss occurred in 10~consecutive epochs, and early stopping was used with patience of 20~epochs. 
{\color{black}A common seed was set to ensure that identical training examples were generated during training across all systems.} {\color{black}For \TCD{} and LRS3, we trained on 3-second segments, whereas 4-second segments were used for WSJ0Mix and LibriMix. This also holds for the length of the reference signal for the \SEA{} systems.}  {\color{black}As the loss function $\bm{\ell}$, the negative \acf{SI-SDR} \cite{leroux2019} was used.}  

\begin{table*}[t!]
\centering
\setlength{\tabcolsep}{1.2em}
\renewcommand{\arraystretch}{1.4}
\caption{Extraction performance of the systems for fully-overlapped 2-speaker and 3-speaker mixtures.}
\sisetup{table-format = 2.1}
\begin{tabular}{@{}llcccccccccccc@{}}
\toprule
     &         &  & \multicolumn{2}{c}{\textbf{TCD-TIMIT}}                   &  & \multicolumn{2}{c}{\textbf{LRS3}}                        &  & \multicolumn{2}{c}{\textbf{WSJ0Mix}}                        &  & \multicolumn{2}{r}{\textbf{LibriMix}}                 \\ \midrule
     &         &  & \textbf{2}                      & \textbf{3}                      &  & \textbf{2}                      & \textbf{3}                      &  & \textbf{2}                      & \textbf{3}                      &  & \textbf{2}                      & \textbf{3}                      \\ \midrule
SpEx+ \cite{ge2020spex+}& $\Delta$SI-SDR  &  & 15.9           & 10.6        &  & 12.8           & 11.2           &  & 16.1          & 12.2          &  & 14.4          & 11.7          \\
            & ESTOI           &  & 81.6           & 59.4        &  & 77.3           & 62.2            &  & 91.3          & 78.3          &  & 85.4          & 70.3        \\ \midrule
USEV \cite{pan2022usev} & $\Delta$SI-SDR  &  & 18.4           & 14.4        &  & 14.7           & 13.5           &  & -   & -   &  & -   & -          \\
            & ESTOI           &  & 86.2           & 69.0        &  & 80.8           & 69.5            &  & -   & -   &  & -   & -         \\ \bottomrule
SE-A        & $\Delta$SI-SDR  &  & 15.5 & 10.0 &  & 13.7 & 12.0 &  & 15.2 & 13.8 &  & 14.7 & 12.7 \\
            & ESTOI           &  & 81.3 & 60.1 &  & 79.1 & 65.0 &  & 89.1 & 81.4 &  & 85.2 & 73.0 \\ \midrule

SE-V        & $\Delta$SI-SDR  &  & 18.4 & 16.5 &  & 14.4 & 13.2 &  & - & - &  & - & - \\
            & ESTOI           &  & 86.1 & 74.7 &  & 80.4 & 68.5 &  & -   & -   &  & -   & -   \\ \bottomrule
SS+SV       & $\Delta$SI-SDR  &  & 17.9 & 15.4 &  & 13.7 & 10.6 &  & 17.0 & 13.8 &  & 16.4 & 12.9 \\
            & ESTOI           &  & 85.0 & 70.2 &  & 79.2 & 61.6 &  & 92.8 & 80.9 &  & 88.6 & 72.6  \\ \midrule
SS+Oracle   & $\Delta$SI-SDR  &  & 18.0 & 16.1 &  & 14.2 & 11.9 &  & 17.0 & 14.0 &  & 16.5 & 13.7 \\
            & ESTOI           &  & 85.2 & 70.9 &  & 79.8 & 63.4 &  & 92.8 & 81.2 &  & 88.8 & 74.0 \\ \bottomrule
\end{tabular}
\label{tab: overlapped_mixtures}
\end{table*}

\subsection{Evaluation Protocol}
\label{sec: evaluation_protocol}
{ \color{black}
The goal of this study is to gain a better understanding of the role of the auxiliary information in \ac{SE} systems. One important aspect is whether the auxiliary information improves the extraction performance compared to uninformed \ac{SS} systems. We follow prior works~\cite{wang2018voicefilter, vzmolikova2019speakerbeam, xu2020spex, Delcroix2020} that compare the estimated target signal of \ac{SE} with the signal corresponding to the target speaker at the output of \ac{SS}. In particular, oracle selection is applied to select the target in \ac{SS} by computing a reconstruction metric, e.g., \ac{SI-SDR}, between each output and the ground-truth target signal. The signal that yields the maximum value is selected as the target estimate of the \ac{SS} system. We refer to this system as SS+Oracle.

{\color{black} In addition, we provide the results of combining \ac{SS} and \acf{SV} for target speaker selection (SS+SV). This should give some perspective on the practical performance of the SS system when evaluated in a \ac{SE} setup. The \ac{SV} system accepts two utterances as input and computes a similarity score between them. It consists of a speaker embedding module that extracts an embedding vector for each utterance. The similarity is then determined by the cosine distance between the two embeddings. Evaluating target speaker selection for \ac{SS} using \ac{SV} is carried out by computing a similarity score between the enrollment utterance and each output of the \ac{SS} system. The output signal that yields the maximum similarity score is selected as the estimated target speaker. For the speaker embedding module, we employed the state-of-the-art \mbox{ECAPA-TDNN}~\cite{desplanques2020ecapa}, which is pre-trained on the VoxCeleb~1+2 datasets \cite{nagrani2020voxceleb} comprising utterances from over seven thousand speakers. 
We used the implementation provided in SpeechBrain\footnote{\url{https://huggingface.co/speechbrain/spkrec-ecapa-voxceleb}} \cite{SB2021}.}

As evaluation metrics, we use the \ac{SI-SDR} \cite{leroux2019} for speech quality and the \ac{ESTOI} \cite{taal2010short} for speech intelligibility. Audio examples are available online\footnote{\url{https://www.audiolabs-erlangen.de/resources/2023-New-Insights-on-Target-Speaker-Extraction}}.


}


\section{Experimental Results}
\label{sec: experimental_results}
In this study, we investigate the role of the auxiliary information in \ac{SE}. This is first achieved by comparing the performance of the \SEA{} and \SEV{} systems to \ac{SS} {\color{black} for different input conditions}. Then, the behavior of the \ac{SE} systems is examined for different and distorted auxiliary signals given the same mixture signal.


\subsection{Initial Analysis of Fully Overlapped Mixtures}
\label{sec: overlapped_utterances}

{\color{black}In this experiment, we evaluate the extraction performance of the different systems on fully overlapped \mbox{2-speaker} and \mbox{3-speaker} mixtures. We first compare the \SEA{} and \SEV{} systems with the baselines described in Section~\ref{subsec:baselines}. Then, a comparison between the \ac{SE} systems and \ac{SS} is provided. The mean results in terms of the \ac{SI-SDR} improvement ($\Delta$\ac{SI-SDR}) and \ac{ESTOI} are provided in Table~\ref{tab: overlapped_mixtures}. In general, it can be seen that the \ac{ESTOI} scores follow the same trends as the \ac{SI-SDR}.


\subsubsection{Comparison with SE Baselines}
\label{sec: se_baselines}
By comparing the \SEA{} system to SpEx+~\cite{ge2020spex+}, we observe no clear trend with respect to the superiority of either system across the different datasets. 
When examining the performance of \SEV{} and USEV~\cite{pan2022usev}, both systems yield comparable results, except for TCD-TIMIT on \mbox{3-speaker} mixtures, where \SEV{} achieves better performance than the baseline. These results affirm that the adopted \SEA{} and \SEV{} systems are, to a certain extent, competitive with existing \ac{SE} methods in the literature.

\begin{figure*}[t]
\centering
\vspace{1em}
\includegraphics[width=0.99\textwidth]{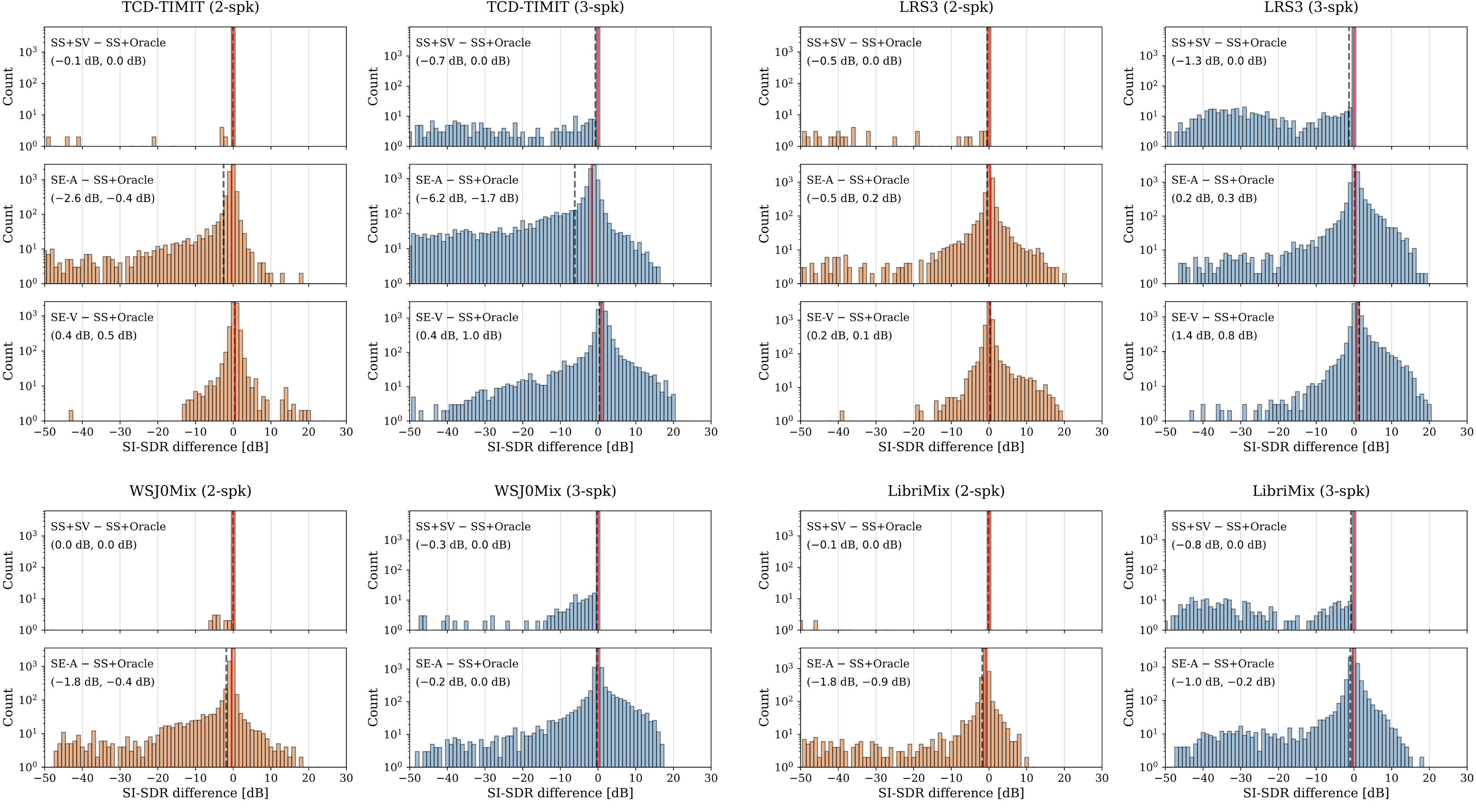}
\caption{{\color{black}Histogram of the SI-SDR [dB] score difference between the different systems and SS+Oracle for \mbox{2-speaker} (orange) and \mbox{3-speaker} (blue) mixtures. The mean (dashed line, black) and median (solid line, red) are also visualized. The tuples provide the values of the mean and median, respectively.}}
\label{fig: diff}
\end{figure*}

\subsubsection{Comparison with SS}
We further compare the performance of the \ac{SE} systems with \ac{SS}+SV and SS+Oracle. We first observe that SS+SV generally yields comparable performance to SS+Oracle for \mbox{2-speaker} mixtures. Conversely, for \mbox{3-speaker} mixtures, SS+SV generally exhibits worse mean performance than SS+Oracle. This can be attributed to the lower separation scores for \mbox{3-speaker} mixtures and the fact that the SV system is pre-trained on clean speech signals rather than on separated signals that might have processing artifacts or residuals from the interfering speakers. Comparing \SEA{} with SS+Oracle, it can be seen that SS+Oracle achieves comparable or better scores across the different datasets. This also holds for SS+SV, except for LRS3 on \mbox{3-speaker} mixtures, where a 1.4~dB drop in performance can be seen compared to \SEA{}. By examining the results of SE-V and SS+Oracle, we observe comparable performance, except for LRS3 on \mbox{3-speaker} mixtures, where \SEV{} achieves better mean scores than SS+Oracle. However, the gap in performance increases when SV is used instead of oracle selection for SS. The results also clearly show that the \SEV{} system exhibits better mean performance than \SEA{}. This could be attributed to the prominent correlation between the visual cues and the target signal in the mixture, as well as the use of time-varying embeddings in \SEV{}.
}

{\color{black}
\subsection{Further Analysis of Fully Overlapped Mixtures}
\label{sec: analysis}
The comparison in Section~\ref{sec: overlapped_utterances} provides a holistic view of the systems' performance, quantified by the arithmetic mean of the evaluation metrics over all samples in the test set. To gain more insights, we perform a per-sample analysis by inspecting the difference in the \ac{SI-SDR} performance between the different systems and SS+Oracle. A negative difference means that the SS+Oracle system is better than the respective system, and vice versa. The histograms of the \ac{SI-SDR} difference are shown in Figure~\ref{fig: diff}. The mean and median values of the \ac{SI-SDR} difference are also provided.  Interestingly, in many cases, the mean and median scores are quite different, which clearly shows that solely reporting the mean performance does not provide a full picture when comparing \ac{SE} and \ac{SS} systems.

Comparing SS+SV with SS+Oracle, it can be observed that the median is always centered around 0~dB. In contrast, the mean deviates to the negative side, especially for \mbox{3-speaker} mixtures, where the SV system sometimes selects an interferer speaker instead of the target. When comparing \SEA{} with SS+Oracle, we observe that, in most cases, the median is close to 0~dB, except for LibriMix on \mbox{2-speaker} mixtures and \TCD{} on \mbox{3-speaker} mixtures, where clearly the \SEA{} system performs poorly compared to SS+Oracle. It is also important to note how susceptible the mean performance of \SEA{} is to outliers, reflected by the gap between the mean and median values. The distributions of the performance difference between \SEV{} and SS+Oracle for \mbox{2-speaker} mixtures exhibit a slight shift towards the positive side for \TCD{}, whereas it is centered around zero for LRS3. However, for \mbox{3-speaker} mixtures, the shift towards the positive side is more prominent, indicating an overall advantage of the visual information in this case. 

The analysis here highlights the inadequacy of reporting only the mean scores over the samples when attempting to compare \ac{SS} and \ac{SE} systems due to the presence of outliers, mainly caused by the incorrect selection of the target speaker. By excluding such outliers and considering the median values, the following conclusions can be drawn. The auxiliary information in \SEA{} does not consistently improve the quality of the extracted signals compared to \ac{SS}, neither for \mbox{2-speaker} mixtures nor for \mbox{3-speaker} mixtures. To some extent, this is also the case for the auxiliary information in \SEV{} for \mbox{2-speaker} mixtures. However, for \mbox{3-speaker} mixtures, the visual information provides an overall improvement compared to \ac{SS}, indicated by the shift of the distributions towards the positive side in Figure~\ref{fig: diff}.
}


\begin{figure*}[t]
\centering
\vspace{1em}
\includegraphics[width=0.99\textwidth]{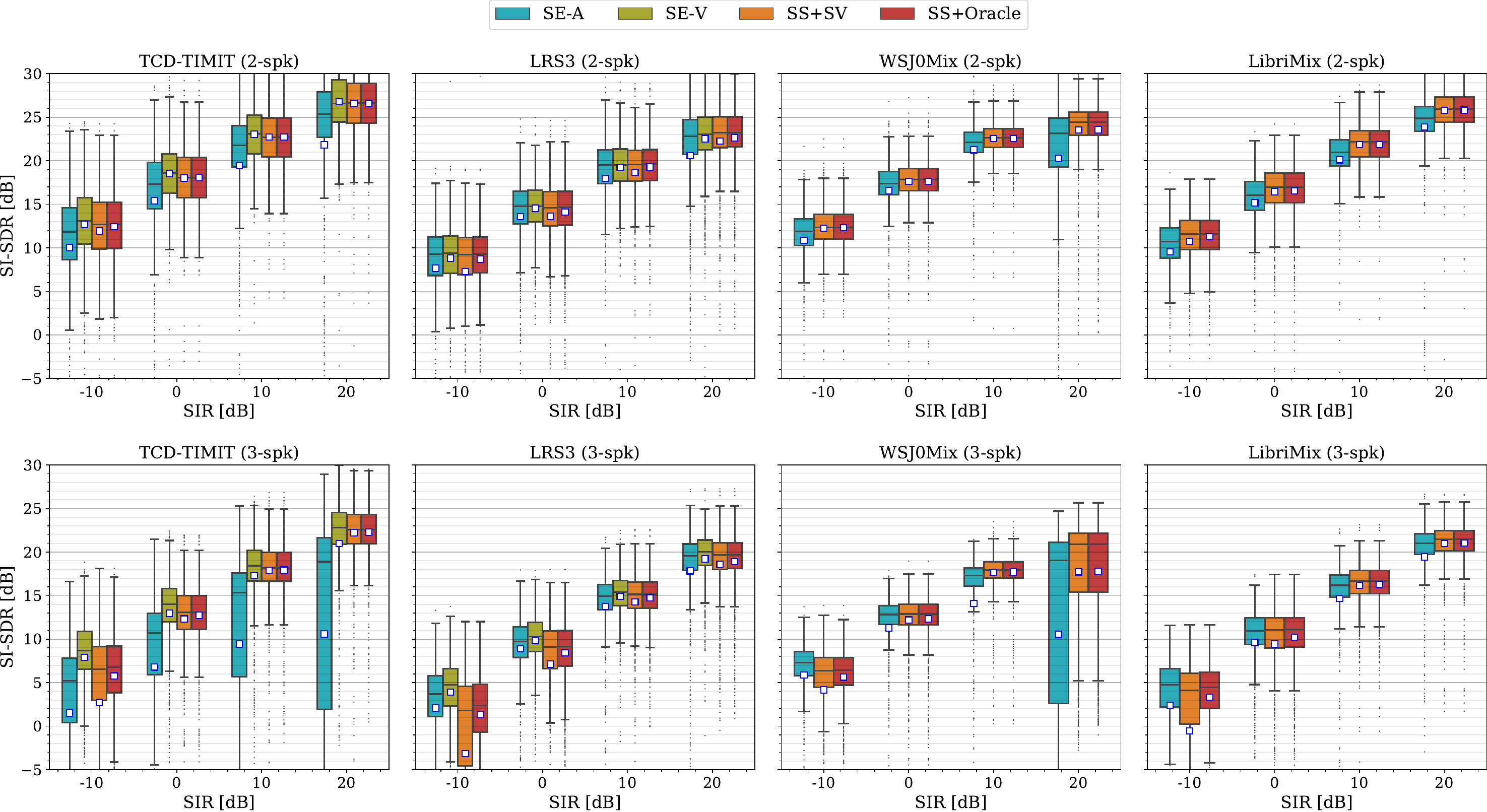}
\caption{Performance in terms of \ac{SI-SDR} [dB] for different \acfp{SIR} for \mbox{2-speaker} (top) and \mbox{3-speaker} (bottom) mixtures. The mean values are visualized by the square symbol.}
\label{fig: SIR}
\end{figure*}

\subsection{Effect of Input SIR}
\label{sec: effect_SIR}
{\color{black}The evaluation in Sections~\ref{sec: overlapped_utterances}~and~\ref{sec: analysis} is carried out on mixtures with varying \ac{SIR}. In this experiment, we specifically study the impact of the input \ac{SIR} on the performance of the different systems for \mbox{2-speaker} and \mbox{3-speaker} mixtures. This examines whether the auxiliary information in \SEA{} and \SEV{} improves the extraction performance compared to \ac{SS} for different powers of the interfering signal(s), especially at low \acp{SIR}.} For evaluation, 1000~examples (target + interferer(s)) were selected from the test split of each dataset and mixed with a \ac{SIR} swept from $-$10~dB to 20~dB with a step size of 10~dB. We report the \ac{SI-SDR} instead of $\Delta$\ac{SI-SDR} to reflect the reconstruction quality of the extracted signals. Figure~\ref{fig: SIR} shows the results of this experiment. As expected, the \ac{SI-SDR} scores generally drop as the \ac{SIR} decreases. {\color{black}Furthermore, it can be seen that SS+SV is comparable to SS+Oracle for higher \acp{SIR} (i.e., $\ge$ 10~dB). However, as the \ac{SIR} decreases, the SS+SV system generally exhibits worse mean performance, especially for \mbox{3-speaker} mixtures. Nonetheless, it is important to note that the median values of SS+SV and SS+Oracle are still close to each other, highlighting again the influence of the outliers on the mean values. 


For \mbox{2-speaker} mixtures, it can be seen that \SEA{} generally performs worse than SS+SV and SS+Oracle in terms of both mean and median values for all \acp{SIR}. Comparing \SEV{} with the \ac{SS} systems, we observe comparable mean and median values, except for LRS3, where the mean of the SS+SV system is lower than \SEV{} and SS+Oracle. The results for \mbox{3-speaker} mixtures follow different trends than the \mbox{2-speaker} case. With the exception of TCD-TIMIT, it can be observed that \SEA{} yields better mean and median scores than SS+SV (and SS+Oracle for LRS3 and WSJ0Mix) for \ac{SIR} $=-$10~dB. However, for higher \acp{SIR}, SS+SV and SS+Oracle generally outperform \SEA{} across the datasets. We can also observe that the \SEA{} system exhibits poor generalization to unseen \acp{SIR} for \TCD{} and WSJ0Mix. It is clear that \SEV{} generally exhibits the best performance in terms of mean and median scores for \ac{SIR} $\le$~0~dB.

The results of this experiment show that the use of auxiliary information in \ac{SE} does not always lead to better performance compared to \ac{SS} for different \acp{SIR}. The only exception is for \mbox{3-speaker} mixtures at low \acp{SIR}, where \SEA{} and \SEV{} generally exhibit better performance than the \ac{SS} systems.
}

\begin{figure*}[t]
\centering
\vspace{1em}
\includegraphics[width=0.99\textwidth]{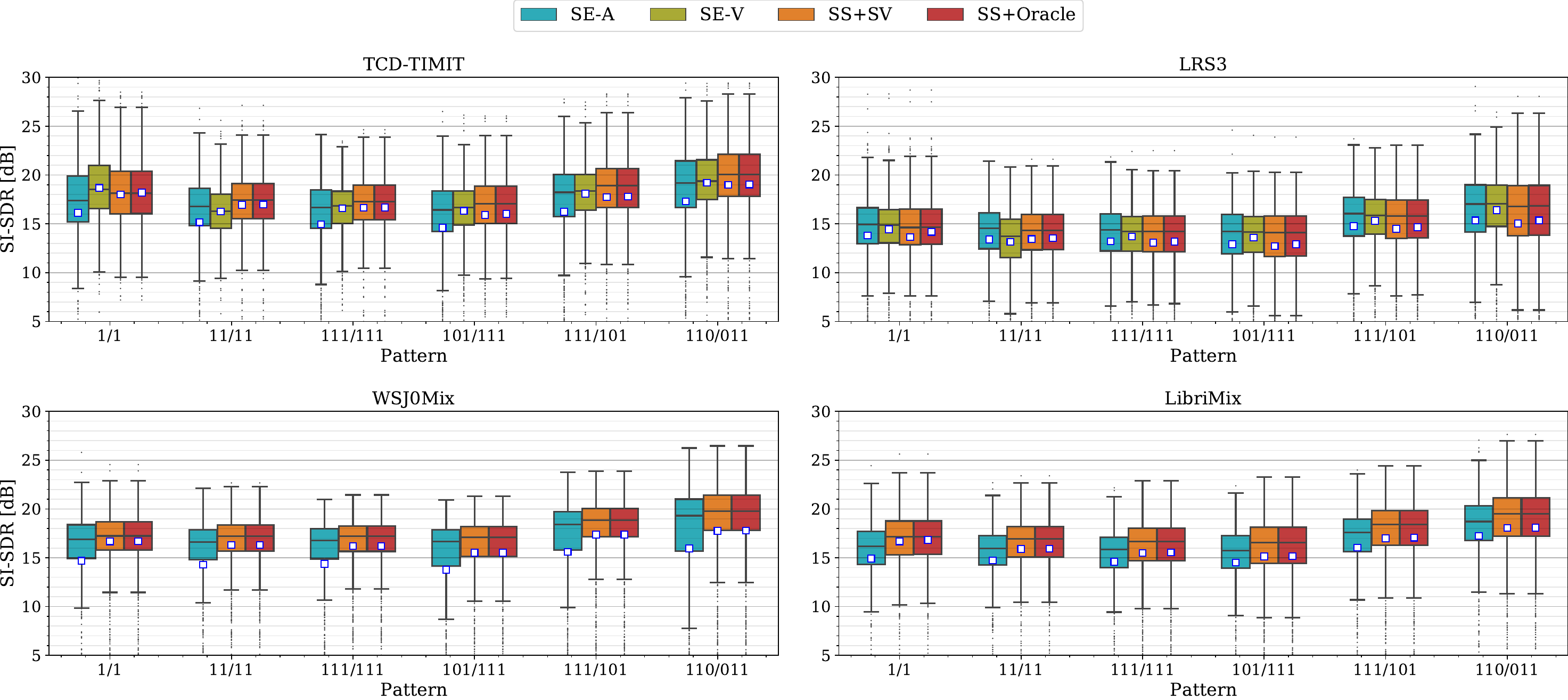} 
\caption{Performance in terms of \ac{SI-SDR} [dB] for the different overlapping patterns for \mbox{2-speaker} mixtures. The mean values are visualized by the square symbol.}
\label{fig: long_seq}
\end{figure*}

\begin{figure}[t]
\centering
\includegraphics[width=0.90\columnwidth]{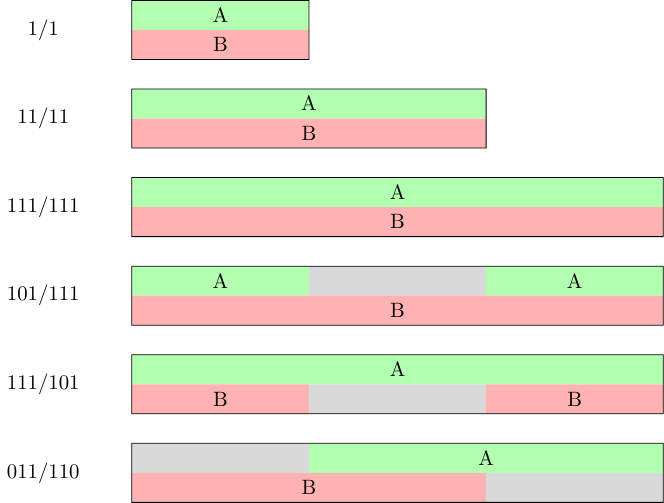} 
\caption{Patterns of the input mixture corresponding to different lengths and different activity patterns of the target speaker~A (green) and the interferer speaker~B (red). Gray regions indicate silence. Each digit (1 or 0) represents a 3-second segment, where 1 denotes a speech segment, whereas 0 represents silence. The last three patterns \{\mbox{101/111}, \mbox{111/101}, \mbox{011/110}\} also include their cyclic shifts by 3~seconds.}
\label{fig: patterns}
\end{figure}

\subsection{Long Sequences}
\label{sec: long_seq}

Thus far, only fully overlapped mixture signals were considered. In this experiment, we evaluate the effectiveness of the auxiliary information in \ac{SE} for \mbox{2-speaker} mixtures with longer durations (up to 9~s) and different overlapping patterns, which mimics what typically occurs in natural conversations. Figure~\ref{fig: patterns} depicts the different overlapping patterns considered in this experiment. For each dataset and each pattern, we generated 1000~mixture signals with a \ac{SIR} of 0~dB. For the \SEV{} system, the duration of the reference signal was always equal to the duration of the mixture signal. {\color{black}In segments where the target speaker was absent, we used static visual frames by repeating the starting frame of the silent segment.} 
For SS+Oracle and SS+SV, target speaker selection was performed on the whole output waveforms rather than over the individual segments. {\color{black}Figure~\ref{fig: long_seq} shows the results of this experiment in terms of \ac{SI-SDR}}. 

Similar to the observations from the previous results on \mbox{2-speaker} mixtures, both SS+Oracle and SS+SV attain comparable mean and median performance across the different patterns and datasets.
Interestingly, the high scores of the \ac{SS} systems indicate that they are able to track the speakers over time, even though there could be pauses in the streams of either the target or interferer, e.g., patterns \mbox{101/111}, \mbox{111/101}, and \mbox{110/011}. This behavior could be attributed to the recurrent structure of the \ac{DPRNN} architecture as well as the non-causality of the \ac{SS} systems. 
{\color{black}From the results, it can be observed that \SEA{} achieves similar performance to SS+SV and SS+Oracle for LRS3, whereas it is generally worse for the other datasets. Note again how different the median and mean values are for \SEA{}, indicating the existence of many outliers where the \SEA{} system performs poorly. A comparison between \SEV{} and SS+SV/Oracle reveals interesting observations. Although, in most cases, the \SEV{} system achieves slightly higher mean values, the median values are generally close to each other or, in some cases, slightly higher for the \ac{SS} systems. Once more, this highlights the importance of reporting not only the mean values when comparing such systems so as to avoid misleading conclusions.}

The findings presented here show that the use of auxiliary information in \ac{SE} does not consistently lead to better performance than \ac{SS} for the considered overlapping patterns for 2-speaker mixtures.




\begin{figure*}[t]
\centering
\includegraphics[width=0.8\textwidth]{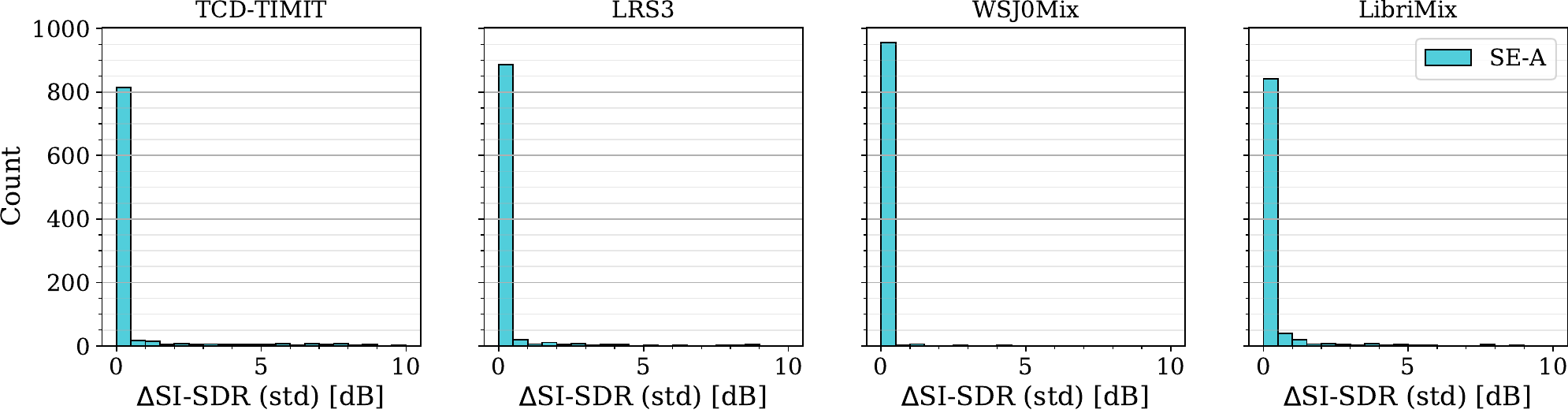}  
\caption{{\color{black}Histograms of the standard deviation in terms of $\Delta$\ac{SI-SDR} [dB] for the \SEA{} system. For each input mixture, the standard deviation of the performance is computed for 5~different randomly selected reference signals belonging to the target speaker.}}
\label{fig: variance_different_adaptations}
\end{figure*}


\begin{figure*}[t]
\centering
\includegraphics[width=\textwidth]{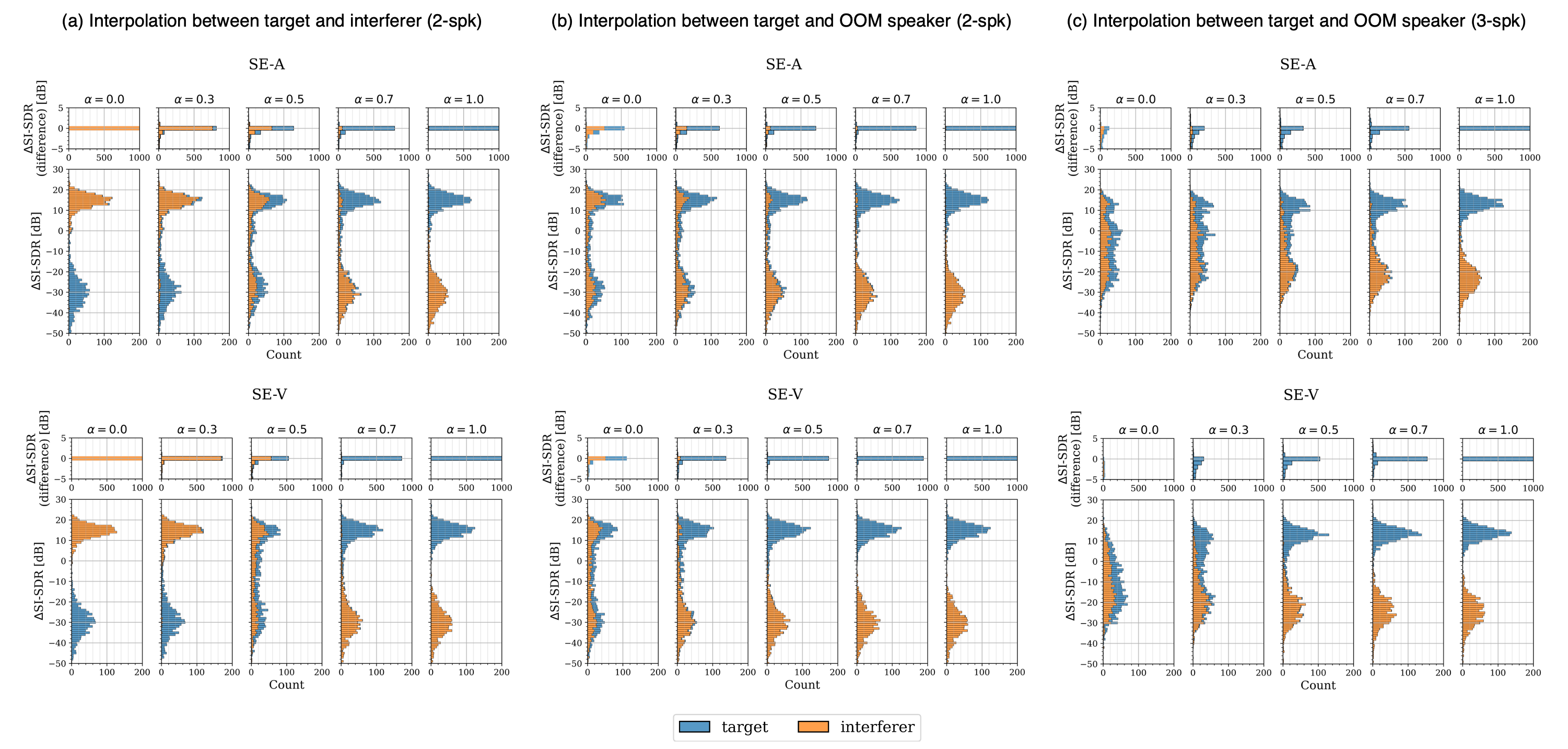}  
\caption{{\color{black} Histograms of the $\Delta$\ac{SI-SDR} scores for each interpolation coefficient $\alpha$.
The corresponding histogram of the difference between the performance of the interpolated embedding and the undistorted embedding for both target ($\alpha=1$), and interferer ($\alpha=0$) is provided on top. (a) Interpolation between the target embedding and the interferer embedding for systems trained and evaluated on \mbox{2-speaker} mixtures; (b) and (c) show the results for interpolation between the target embedding and an out-of-mixture (OOM) speaker embedding for systems trained and evaluated on 2-speaker and 3-speaker mixtures, respectively. The output of the \ac{SE} systems is evaluated with respect to both target (blue) and interferer (orange).}}
\label{fig: interpolation}
\end{figure*}

\subsection{Performance for Different Reference Signals} 
\label{sec: different_reference_signals}

{\color{black}
In the previous experiments, the performance of the \ac{SE} systems was evaluated for different examples, each consisting of a mixture and a reference signal that belongs to the target speaker. In this experiment, however, we investigate whether the extraction performance varies for different auxiliary signals from the target speaker given the same input mixture. Note that this experiment is only carried out for \SEA{}, since different enrollment utterances for the target speaker can be obtained, which are different from the target speech in the mixture. However, for \SEV{}, there is only a unique reference signal corresponding to the visual frames of the spoken utterance in the mixture. For evaluation, 1000~\mbox{2-speaker} mixtures with a \ac{SIR} of 0~dB were considered. For each mixture, we evaluate the extraction performance for 5~different randomly selected reference signals. 
The standard deviation of $\Delta$\ac{SI-SDR} for each mixture is computed, and the histogram for each dataset is depicted in Figure~\ref{fig: variance_different_adaptations}. Interestingly, across the different datasets, in more than 80\% of the cases, the standard deviation is less than 0.5~dB, which demonstrates that the extraction performance for the 5~different reference signals is generally close to each other. 
These results suggest that the quality of the reconstructed target speech does not vary for different enrollment utterances having `sufficient' discriminative information. 
}

\subsection{Distorted Auxiliary Information} 


{\color{black}
We further investigate the role of the auxiliary information on the extraction performance of \ac{SE}} by inspecting the outcome of the \SEA{} and \SEV{} systems for distorted auxiliary information. As a distortion, we linearly interpolate between the target speaker embedding and another embedding, either obtained from the interfering speaker or an out-of-mixture (OOM) speaker, i.e., \mbox{$\bm{E}_\text{interpolated} = \alpha\, \bm{E}_\text{target} + (1-\alpha)\, \bm{E}_{\text{\{interferer, OOM\}}}$}. Although this setup was not part of the training phase, this experiment explores how much \ac{SE} relies on the auxiliary information and how different the extraction performance is compared to the results for undistorted embeddings. We consider both \SEA{} and \SEV{} systems {\color{black}trained independently on \mbox{2-speaker} and \mbox{3-speaker} mixtures}. For evaluation, 1000~\mbox{2-speaker} {\color{black} and \mbox{3-speaker}} mixtures with a \ac{SIR} of 0~dB were considered. We evaluate the reconstruction quality in terms of $\Delta$\ac{SI-SDR} with respect to the target as well as the interferer to determine which speaker was extracted. 
In addition, we report the performance difference between the interpolated embedding and the undistorted embeddings of both the target and interferer. A performance difference close to zero indicates that the interpolated embedding allows extracting the target/interferer speaker with a reconstruction quality close to the respective undistorted embedding. 
The results are provided in Figure~\ref{fig: interpolation}. For brevity, we only show the results for the LRS3 dataset, as the results for the other datasets exhibit a similar pattern. 


For the \ac{SE} systems trained and evaluated on \mbox{2-speaker} mixtures, the results for interpolation with the interferer embedding in \mbox{Figure~\ref{fig: interpolation}a} show that, in general, the closer the interpolated embedding is to either the target or interferer embedding, the more likely the corresponding speaker is extracted. Except for $\alpha=0.5$, it is interesting to observe from the histogram of differences that the performance of both interpolated and undistorted embeddings is, in most cases, close to each other. 
For $\alpha=0.5$, in at least 50\% of the cases, both \SEA{} and \SEV{} systems still equally likely extract either the target or interferer with the same quality as when the undistorted embedding is provided. 

When an embedding from an OOM speaker is interpolated with the target embedding in \mbox{2-speaker} mixtures, it can be seen in \mbox{Figure~\ref{fig: interpolation}b} that for higher values of $\alpha$, the target speaker is more likely to be extracted compared to the interferer. Interestingly, when an utterance from an OOM speaker is used as a reference signal for \SEA{} (i.e., $\alpha=0$), in 50\% of the cases, the system extracts either the target or interferer with the same quality as when the undistorted embedding is provided. A similar behavior can be seen for the \SEV{} system at $\alpha=0$, although the reference signal (lip frames), in this case, corresponds to a different utterance spoken by an OOM speaker. 

{\color{black}For the \ac{SE} systems trained and evaluated on \mbox{3-speaker} mixtures, the results for interpolation with one of the interferer's embeddings are omitted for brevity, as they follow the ones shown in \mbox{Figure~\ref{fig: interpolation}a}. However, when interpolation is performed between the target speaker and an OOM speaker, it can be seen in \mbox{Figure~\ref{fig: interpolation}c} that the \ac{SE} systems tend to fail to extract any speaker in the mixture as $\alpha$ decreases. This behavior is different from that of the systems trained on \mbox{2-speaker} mixtures (in \mbox{Figure~\ref{fig: interpolation}b}), where the systems are inclined to extract either speaker in the mixture. This indicates how influential the training strategy, particularly the training data composition in terms of the number of speakers in the mixture, is to the way the \ac{SE} systems utilize the auxiliary information. 

In this experiment, we inspected the behavior of the \SEA{} and \SEV{} systems for different points in the embedding space given the same mixture signal. For systems trained on \mbox{2-speaker} mixtures, we found that the \ac{SE} systems generally tend to extract either speaker in the mixture, even when the provided embedding is drawn from an OOM speaker. This behavior, however, is different for systems trained on \mbox{3-speaker} mixtures, in which providing an embedding from an OOM speaker does not lead to extracting any of the constituent speakers in the mixture. These results suggest that the way the \ac{SE} systems are trained significantly affects how the auxiliary information is exploited to extract the target speaker.
}

\section{Discussion}
\label{sec: discussion}

In this study, we examined the role of the auxiliary information in two reference-based \ac{SE} systems, namely audio-based \ac{SE} (\SEA{}) and video-based \ac{SE} (\SEV{}). In the first set of experiments, we compared the extraction performance of both \ac{SE} systems to an uninformed \acf{SS} system evaluated in a SE framework. The comparison was carried out for 2-speaker and 3-speaker mixtures across different datasets. {\color{black}We demonstrated that the use of auxiliary information in the \ac{SE} systems does not always result in better extraction performance than \ac{SS}. However, we showed that the auxiliary information could be helpful for 3-speaker mixtures under low \acp{SIR}.} 
{\color{black} In the second set of experiments, we inspected the behavior of \ac{SE} for different auxiliary signals given the same mixture signal. We first showed that the performance of \SEA{} does not generally vary for different enrollment signals from the target speaker for a given mixture. In addition, we evaluated the performance of \ac{SE} systems for distorted auxiliary information by mixing the target speaker embedding with an embedding either from the interferer or a speaker not in the mixture. We showed that distorting the auxiliary information can, in general, either have an insignificant effect on the quality of the reconstructed target signal, confuse the system into extracting the interferer signal with a similar performance to when an auxiliary signal from the interferer is provided, or make the system fail to extract either speaker. We also demonstrated that the training strategy employed for the \ac{SE} systems greatly influences the way the auxiliary information is utilized in \ac{SE}.}


{\color{black} While this study provides valuable insights into the role of auxiliary information in SE, it is crucial to acknowledge its limitations to avoid over-generalization of the results. On the system level, we considered the \ac{DPRNN} architecture as the main learning machine for all systems, although several new architectures \cite{subakan2021attention, wang2022tf} have been proposed and demonstrated better separation performance. Therefore, whether the conclusions reached in this study also extend to these architectures is yet to be validated. In addition, we considered only two forms of auxiliary information for \ac{SE}: enrollment utterances and visual information. It would be interesting to extend this study to other forms of auxiliary information and possibly also their combinations, i.e., multi-modal \ac{SE}. Furthermore, we used the same hyperparameters in all experiments. Ideally, however, there should be a hyperparameter optimization stage for each system and each dataset. 
Finally, the list of input mixture scenarios is by no means exhaustive, e.g., neither non-speech interferers nor reverberation was considered.} 

Nonetheless, the findings in this study suggest that existing \ac{SE} methods do not fully exploit the auxiliary information in extraction, which, intuitively, should significantly improve the performance compared to \ac{SS}. Therefore, one future direction would be to explore ways to better utilize the auxiliary information in \ac{SE}. Furthermore, if the auxiliary information would mainly help to select the speaker of interest, then this implies that {\color{black}as long as sufficient discriminative} information is present in the auxiliary signal, then reasonable extraction performance should be expected. This could drive further works that cope with unreliable auxiliary information, e.g., noisy enrollment utterances from the target speaker or occluded/misaligned visual frames.  



%




\ifCLASSOPTIONcaptionsoff
  \newpage
\fi



\bibliographystyle{IEEEtran}
\bibliography{sapref}

\end{document}